\documentclass{ws-mpla-arXiv11015370}

\usepackage{amsmath,amssymb,graphicx}  

\begin{document}

\markboth{F.R. Klinkhamer}
         {Newton's gravitational coupling constant from a quantum of area}

\catchline{}{}{}{}{}

\title{\vspace*{-10mm}Newton's gravitational coupling constant
       from a quantum of area}

\author{F.R. Klinkhamer}

\address{Institute for Theoretical Physics, University of Karlsruhe,\\
             Karlsruhe Institute of Technology, 76128 Karlsruhe,
             Germany\\
frans.klinkhamer@kit.edu}

\maketitle


\begin{abstract}
A previous calculation of Newton's gravitational coupling constant $G$ is
generalized. This generalization makes it possible to have ``atoms of
two-dimensional space'' with an integer dimension $d_\mathsf{atom}$ of the
internal space, where the case $d_\mathsf{atom}=1$
is found to be excluded. Given the
quantum of area $l^2$, the final formula for $G$ is inversely proportional to
the logarithm of the integer $d_\mathsf{atom}$. The generalization used may
be interpreted as a modification of the energy equipartition law of the
microscopic degrees of freedom responsible for gravity, suggesting some form
of long-range interaction between these degrees of freedom
themselves.
\vspace*{0.5\baselineskip}\newline
Journal: Mod. Phys. Lett. A {\bf 26}, 1301 (2011)
\vspace*{0.5\baselineskip}\newline
Preprint: arXiv:1101.5370v6
\keywords{Fundamental constants; Newtonian mechanics; thermodynamics.}
\end{abstract}
\ccode{PACS numbers: 06.20.Jr, 45.20.D-, 05.70.-a}

\section{Introduction}\label{sec:introduction}

It has been argued\cite{Klinkhamer2007} that the fundamental length scale of
quantum spacetime need not be given by the Planck length, $l_{P}\equiv
(\hbar\, G)^{1/2}/c^{3/2} \approx  1.6 \times 10^{-35}\;\text{m}$, but may
correspond to a new fundamental constant of nature, $l$. This would then
suggest that Newton's gravitational coupling constant $G$ becomes calculable
in terms of the fundamental constants $c$ (velocity of light in vacuum),
$\hbar$ (Planck's quantum of action), and $l$ (the hypothetical quantum of
length).

Stimulus for a calculation of $G$ was provided by the approach of
Verlinde\cite{Verlinde2010} to consider the Newtonian gravitational
attraction as a  type of entropic force, with the fundamental microscopic
degrees of freedom living on a two-dimensional screen, in line with
the so-called holographic principle.\cite{'tHooft1999}
Following this approach
and using the Bekenstein--Hawking black-hole
entropy\cite{Bekenstein1973,Hawking1975,Unruh1976,Hawking1996}
as input, a single
transcendental equation can be derived, which fixes the numerical factor $f$
entering the $G$ expression.\cite{Klinkhamer2010}

Now, it is possible to make further progress by combining two recent
suggestions. The first is by Sahlmann (last paragraph in
Ref.~\refcite{Sahlmann2010}) that the internal Hilbert-space dimension of the
``atom of two-dimensional space,'' corresponding to the ``quantum of area''
$l^2$, may very well need to be integer and that this places further
restrictions on the microscopic theory. The second is by Neto\cite{Neto2011}
that the microscopic degrees of freedom on the holographic screen may have a modified
energy equipartition law. Such a behavior may result from a generalization of
the standard Boltzmann-Gibbs
statistics,\cite{Tsallis1988,Tsallis1999,PlastinoLima1999} but it may also
have an entirely different origin. (The important role of the equipartition
law has previously been emphasized in, e.g., Ref.~\refcite{Padmanabhan2009}.)

Prompted by these two suggestions, a new calculation of $G$
is presented in this paper.
In addition, a physical interpretation of the result can be given,
which is based on the Verlinde approach to the origin of gravity.

\section{Combinatorial calculation}
\label{sec:Combinatorial-calculation}

This section gives a purely combinatorial calculation
of the numerical factor $f$ entering the general expression for the gravitational coupling
constant:
\begin{equation}\label{eq:G-Ansatz}
G=f\,c^3\,l^2/\hbar\,,
\end{equation}
where $l^2$ is considered to be a new fundamental constant of nature with the
dimension of area. Incidentally, the notation $G_{N}$ will be kept for the
experimental value of Newton's gravitational coupling
constant.\cite{MohrTaylorNewell2008}

Following Sec.~4 of Ref.~\refcite{Klinkhamer2010}, there are two steps for
the combinatorial calculation of $f$. First, consider the event horizon of a
large nonrotating (Schwarzschild) black-hole
and write the number of degrees of freedom on this
surface (with area $A$) as the product of two dimensionless numbers,
\begin{equation}\label{eq:Ndof-atom}
N_\mathsf{dof} = d_\mathsf{atom}\,N_\mathsf{atom}\,.
\end{equation}
Here, $N_\mathsf{atom}$ is interpreted as the number of distinguishable
``atoms of two-dimensional space'' making up the area ($l^2$ being the
quantum of area) and $d_\mathsf{atom}$ as the dimension of the internal space
of an individual atom:
\begin{subequations}\label{eq:N-atom-d-atom}
\begin{eqnarray}\label{eq:N-atom}
N_\mathsf{atom}&\equiv& A/l^2
                \in \mathbb{N}_1 \equiv \{1,\, 2,\, 3,\, \ldots\}\,,
\\[1mm]
\label{eq:d-atom} d_\mathsf{atom} &\equiv& f^{-1}\,I_{1}^{-1} \in
\mathbb{N}_1\,.
\end{eqnarray}
\end{subequations}
Compared to the analysis of Ref.~\refcite{Klinkhamer2010}, there are two new
ingredients in \eqref{eq:d-atom}: $d_\mathsf{atom}$ is no longer equal to
$f^{-1}$ and $d_\mathsf{atom}$ is demanded to be a positive integer.
The factor $I_{1}^{-1}$ in definition \eqref{eq:d-atom} simply
parameterizes the difference of $d_\mathsf{atom}$ and $f^{-1}$.
For the moment, the origin and meaning of $I_{1}^{-1}$ is left open
(one possible physical interpretation will
be given in Sec.~\ref{sec:Modified-equipartition-law}). From now on,
abbreviate ``atoms of two-dimensional space'' as ``atoms of space'' or even
``atoms.'' One such ``atom'' will be said to contribute one ``quantum of
area'' $l^2$ to a macroscopic surface.

Second, take as input the Bekenstein--Hawking
formula\cite{Bekenstein1973,Hawking1975} for the entropy of a large
(macroscopic) black-hole
\begin{equation}\label{eq:S-BH}
S_\mathsf{BH}/k_{B}
= c^3 A/(4\,\hbar\,G) =
(1/4)\,f^{-1}\,A/l^2 =
(1/4)\,I_{1}\,d_\mathsf{atom}\,N_\mathsf{atom} \,,
\end{equation}
where \eqref{eq:G-Ansatz} has been used in the  second step and
\eqref{eq:N-atom} and  \eqref{eq:d-atom} in the third step.
Equating the
number of configurations of the distinguishable atoms of space from
\eqref{eq:Ndof-atom} with the exponential of the Bekenstein--Hawking entropy
\eqref{eq:S-BH} gives the following set of conditions:
\begin{equation}\label{eq:Natom-datom-condition}
\big(d_\mathsf{atom}\big)^{\,N_\mathsf{atom}} =
\exp\Big[(1/4)\,I_{1}\,d_\mathsf{atom}\,N_\mathsf{atom}\Big]\,,
\end{equation}
for positive integers $N_\mathsf{atom}\gg 1$ (there may be significant
corrections to the black-hole entropy for $N_\mathsf{atom}\sim 1$; see, e.g.,
Ref.~\refcite{Engle-etal2010} and references therein).
The infinite set of conditions \eqref{eq:Natom-datom-condition}
reduces,  for given $I_{1}$, to a single transcendental equation
for $d_\mathsf{atom}$,
\begin{subequations}\label{eq:datom-transcendental-eq-integer}
\begin{eqnarray}\label{eq:datom-transcendental-eq}
\ln\,d_\mathsf{atom} &=& (1/4)\,I_{1}\,d_\mathsf{atom}\,.
\end{eqnarray}
In addition, there is still the condition that the dimension of the internal
space be a positive integer,\cite{Sahlmann2010}
\begin{eqnarray}
\label{eq:datom-integer} d_\mathsf{atom} &\in&\ \mathbb{N}_1\,.
\end{eqnarray}
\end{subequations}
Note that \eqref{eq:datom-transcendental-eq} has precisely the same form as
Eq.~(13) of Ref.~\refcite{Klinkhamer2010}, except for the additional factor
$I_{1}$ on the right-hand side. Similar modifications can be expected for the
generalized models of Ref.~\refcite{Sahlmann2010}.

Table~\ref{table1} gives the required $I_{1}$ values (indicated by hats) from
\eqref{eq:datom-transcendental-eq} to make for integer $d_\mathsf{atom}$
values. Three remarks are in order. First, having a solution of
\eqref{eq:datom-transcendental-eq} demands a small enough numerical factor
$(1/4)\,I_{1}$ on the right-hand side, corresponding to $I_{1} \leq 4/e
\approx 1.47152$, with $e\approx 2.71828$ the base of the natural logarithm.
Second, the required $I_{1}$ values for $d_\mathsf{atom}=2$ and
$d_\mathsf{atom}=4$ are equal, but it is not clear if this carries over to
generalized models (for example, those of Ref.~\refcite{Sahlmann2010}).
Third, the value $d_\mathsf{atom}=1$ is physically not allowed, as $I_{1}=0$
from \eqref{eq:datom-transcendental-eq} implies a vanishing black-hole
entropy \eqref{eq:S-BH} for $d_\mathsf{atom}=1$ and finite $N_\mathsf{atom}$.

\begin{table}[t]  
\renewcommand{\tabcolsep}   {4pc}    
\renewcommand{\arraystretch}{1.125}   
\tbl{ Selected $I_{1}$ values required for having integer $d_\mathsf{atom}$
values, according to \eqref{eq:datom-transcendental-eq} and
\eqref{eq:datom-integer}. Also shown is the corresponding $q$ value from
\eqref{eq:I1-2dMaxwell}. The value $d_\mathsf{atom}=1$ is nonphysical,
because $\widehat{I}_{1}$ is found to vanish. If \eqref{eq:I1-2dMaxwell}
holds, the values $d_\mathsf{atom}\geq 27$ are, most likely, also
nonphysical, because the values $\widehat{q}$ turn out to be
negative.\protect\cite{Tsallis1988}
} {\begin{tabular}{@{}ccc@{}} \toprule $d_\mathsf{atom}=\widehat{d}\equiv n$&
$\widehat{I}_{1}\equiv (4\,\ln n)/n$&
$\widehat{q}\equiv 2- 1/\widehat{I}_{1}$ \\
\colrule
(1)& (0)       & ($-\infty$) \\
2  & 1.3863    & 1.2787 \\
3  & 1.4648    & 1.3173 \\
4  & 1.3863    & 1.2787 \\
5  & 1.2876    & 1.2233 \\
8  & 1.0397    & 1.0382 \\
9  & 0.9765    & 0.9760 \\
26 & 0.5012    & 0.0050 \\
27 & 0.4883 &  $-0.0480\phantom{0.}$\\
\botrule
\end{tabular}
\label{table1}}
\end{table}

With the solutions of \eqref{eq:datom-transcendental-eq} and
\eqref{eq:datom-integer}, the final formula for  Newton's gravitational
coupling constant $G$ from \eqref{eq:G-Ansatz} reads
\begin{subequations}\label{eq:G-final-dhat}
\begin{eqnarray}\label{eq:G-final}
G &=& \big(1/4\big)\,\big(\ln\,\widehat{d}\;\,\big)^{-1}\;
       c^3\,l^2/\hbar\,,
\\[1mm]
\widehat{d} &\in& \mathbb{N}_1\backslash \{1\}
              =   \{2,\,3,\,4,\, \ldots\,\}\,,
\label{eq:dhat}
\end{eqnarray}
\end{subequations}
where $\widehat{d}$ is the internal dimension of an atom of space with
quantum of area $l^2$. The fundamental microscopic theory will have to decide
which value of $\widehat{d}$ appears in \eqref{eq:G-final}. Observe that,
given $l^2$, the maximal value of $G$ is obtained for the minimal value of
the integer $\widehat{d}$, namely, $\widehat{d}=2$.

 From the experimental value
$G_{N}= 6.6743(7)\;10^{-11}\;\text{m}^{3}\;\text{kg}^{-1}\;\text{s}^{-2}$
(see, e.g., Chap. X of Ref.~\refcite{MohrTaylorNewell2008} for further
discussion), the following numerical estimate of the smallest possible
quantum of area is obtained:
\begin{eqnarray}\label{eq:l2-num}
l^2\;\Big|_{\,\widehat{d}=2} &=&
4\,\ln 2\; \big(l_{P}\big)^2 \approx
7.2423 \times 10^{-70}\;\text{m}^2 \,,      
\end{eqnarray}
with $l_{P}\equiv (\hbar\, G_{N})^{1/2}/c^{3/2} \approx  1.6162 \times
10^{-35}\;\text{m}$. Not surprisingly, this particular value of the quantum
of area has already been given in an earlier article by 't
Hooft.\cite{'tHooft1999}
Here, there is the further result that the dimension
$d_\mathsf{atom}=\widehat{d}=1$ is
ruled out on physics grounds, leaving $\widehat{d}=2$
as the lowest
possible value and allowing for the storage of information (``bits'') on
a holographic screen.\cite{'tHooft1999}

Expression \eqref{eq:G-final-dhat} for $G$ is the main result of
this paper.
The crucial Eqs.~\eqref{eq:datom-transcendental-eq} and \eqref{eq:datom-integer}
for its derivation rely only on the definitions
\eqref{eq:G-Ansatz}--\eqref{eq:N-atom-d-atom},
the interpretation of $N_\mathsf{dof}$ mentioned in the lines under
\eqref{eq:Ndof-atom}, and the input \eqref{eq:S-BH} corresponding to
the entropy of a Schwarzschild black hole.\footnote{The same input
is obtained from a de-Sitter universe (Hubble constant $H$).
Setting $G=\hbar=c=k_{B}=1$, this spacetime has, in fact,
a Hawking temperature $T=H/(2\pi)$,
an event-horizon area $A=4\pi/H^2$,
and an entropy $S=\pi/H^2=(1/4)\,A$;
see, e.g., Chap. 5, p. 88 of Ref.~\refcite{Hawking1996}.
}
The really new
ingredient, here, is the extra factor $I_{1}^{-1}$ in definition
\eqref{eq:d-atom}. The rest of this paper
is devoted to one possible
physical interpretation of $I_{1}$, but there may, of course, be other
interpretations.

\section{Modified equipartition law and entropic gravity}
\label{sec:Modified-equipartition-law}

As mentioned in the Introduction, it has been suggested\cite{Neto2011} that
the microscopic degrees of freedom responsible for gravity obey a modified
equipartition law which can be written as\cite{PlastinoLima1999}:
\begin{equation}\label{eq:E-equipartition}
E = N_\mathsf{dof}\;\frac{1}{2}\;I_{1}\;k_{B}T\,,
\end{equation}
where, for the moment, the real factor $I_{1}>0$ is considered to be
unrelated to the
quantity $I_{1}^{-1}$ appearing in \eqref{eq:d-atom}. At first, it may be
best to remain agnostic as to the possible origin
of the nonstandard equipartition law \eqref{eq:E-equipartition}
with $I_{1}\ne 1$.
A particular calculation of $I_{1}\ne 1$
from nonstandard statistics\cite{Tsallis1988,Tsallis1999,PlastinoLima1999}
will, however, be considered in Sec.~\ref{sec:Generalized-statistics}.

Taking \eqref{eq:E-equipartition} for granted, return to the derivation (4)
in Ref.~\refcite{Klinkhamer2010} of the Newtonian gravitational acceleration
$\mathbf{A}_\mathsf{grav}$ on a test mass arising from a spherical
holographic screen $\Sigma_\mathsf{sph}$ with area $A=4\pi R^2$:
\begin{eqnarray}\label{eq:Agrav-derivation}
\big|\mathbf{A}_\mathsf{grav}\big|
&\stackrel{\mathsf{\textcircled{\sffamily\tiny 1}}}{=}& 2\pi\,   c\,  \big(
k_{B}T /\hbar\big)
\nonumber\\[1mm]
&\stackrel{\mathsf{\textcircled{\sffamily\tiny 2'}}}{=}&
4\pi\,  f c \,
\big(N_\mathsf{dof}\,\textstyle{\frac{1}{2}}\,I_{1}\,  k_{B}T/\hbar\big) \,
\big(f^{-1}\,I_{1}^{-1} / N_\mathsf{dof}\big)
\nonumber\\[1mm] &\stackrel{\mathsf{\textcircled{\sffamily\tiny 3'}}}{=}&
4\pi\,  f c \, \big(E          /\hbar\big)\,  \big(l^2 / A\big) \nonumber\\
[1mm] &\stackrel{\mathsf{\textcircled{\sffamily\tiny 4}}}{=}&
f   \,c\, \big(M c^2      /\hbar\big) \, \big(l^2 / R^2\big)\nonumber\\
[1mm] &\stackrel{\mathsf{\textcircled{\sffamily\tiny 5}}}{=}& \big(f \,c^3 \,
l^2/\hbar\big) \;  M/R^2\,.
\end{eqnarray}
Step $1$ in the above derivation relies on the Unruh
temperature\cite{Unruh1976} (but with the logic reversed, temperature giving
rise to acceleration\cite{Verlinde2010}). Step $2'$ uses straightforward
mathematics and prepares the way for the next move. Step $3'$, then, relies
on \eqref{eq:E-equipartition} and the following relation between the number
$N_\mathsf{dof}$ of degrees of freedom on the holographic screen and the area
$A$ of the screen:
\begin{equation}\label{eq:Ndof}
 N_\mathsf{dof} = f^{-1}\,I_{1}^{-1}\, A/l^2\,.
\end{equation}
Step $4$ depends on the well-known relation of energy and mass from special
relativity. Step $5$, finally, separates the fundamental microscopic
constants of nature (indicated by lower-case letters) from the macroscopic
variables of the experimental setup (upper-case letters).

The last expression in \eqref{eq:Agrav-derivation} gives the Newtonian
gravitational coupling constant $G$ in the form \eqref{eq:G-Ansatz}.
That is, the classical constant $G$ is obtained as a
\textit{ratio} of the two quantum
constants $l^2$ and $\hbar$ (this point has already
been emphasized in Ref.~\refcite{Klinkhamer2007}). Furthermore,
\eqref{eq:Ndof} concords with the previous definitions \eqref{eq:Ndof-atom},
\eqref{eq:N-atom}, and \eqref{eq:d-atom}. The derivation
\eqref{eq:Agrav-derivation} identifies, therefore, the number $I_{1}^{-1}$
entering definition \eqref{eq:d-atom} as the inverse of the modification
factor $I_{1}$ of the equipartition law \eqref{eq:E-equipartition}.
The numerical values of $I_{1}$ and $f$ are determined by
the calculation of Sec.~\ref{sec:Combinatorial-calculation}.

Hence, the suggestion is that the atoms of space
have some type of long-range
interaction or long-time memory, which results in a modification of the
energy equipartition law \eqref{eq:E-equipartition}. The numerical values for
$I_{1}$ in Table~\ref{table1} show that, for the simplest atom with
$\widehat{d}=2$, the standard equipartition law must be modified by
some $+40\,\%$. Note also that the
$\widehat{I}_{1}$ values in Table~\ref{table1}
are larger than 1 for dimensions
$2\leq \widehat{d}\leq 8$ and smaller than 1 for $\widehat{d}\geq 9$.

\section{Generalized statistics}
\label{sec:Generalized-statistics}

Now, consider one possible explanation of the nonstandard equipartition
law \eqref{eq:E-equipartition},
namely, the generalization of the standard Boltzmann-Gibbs statistics along
the lines suggested by
Tsallis.\cite{Tsallis1988,Tsallis1999}
This allows for an explicit calculation of the modification factor $I_{1}$ in
the equipartition law \eqref{eq:E-equipartition}, as a function of the
nonextensive entropy index $q\in \mathbb{R}$ of
Tsallis.\cite{Tsallis1988}

For a quadratic classical Hamiltonian, the modified equipartition law has
been derived in Eq.~(32) of Ref.~\refcite{PlastinoLima1999}, with $I_{1}$
defined by Eq.~(47) of that same reference. Specifically, for a generalized
Maxwell velocity distribution in two-dimensional Euclidean space, the
following result holds\cite{PlastinoLima1999}:
\begin{equation}\label{eq:I1-2dMaxwell}
I_{1} = \frac{\int_{0}^{1} du \,u\; [1-u^2]^{1/(1-q)}}
             {\int_{0}^{1} du \,u\; [1-u^2]^{q/(1-q)}}
      = \frac{1}{2-q}\;,
\end{equation}
for $0<q<2$. The standard Boltzmann--Gibbs statistics ($q=1$)
gives $I_{1}=1$.

The index $q$ enters the generalized entropy relation for two independent
systems, $L$ and $R$, in the following way\cite{Tsallis1988,Tsallis1999}:
\begin{equation}\label{eq:s-nonextensive}
s(L+R)/k_q= s(L)/k_q+s(R)/k_q+(1-q)\;s(L)/k_q\;s(R)/k_q\,,
\end{equation}
where, for clarity, the nonstandard entropy is denoted by a lower-case letter
`$s$' and $k_q$ is a new Boltzmann-type constant with the only requirement
that $k_1 = k_{B}$. From the numerical values for $q$ in Table~\ref{table1},
a system of atoms of space with internal dimensions $2\leq \widehat{d}\leq 8$
then has a \emph{subadditive} entropy $s$ and a system of atoms with
dimensions $9 \leq \widehat{d}\leq 26$ a \emph{superadditive} entropy $s$.
Systems of atoms with $\widehat{d}\geq 27$ have unusual
(most likely, unacceptable) thermodynamics, with, for example,
a convex entropy.\cite{Tsallis1999}

If the modified equipartition law \eqref{eq:E-equipartition} is indeed due to
a form of nonstandard statistics as suggested in the previous paragraph, then
the following question arises:
how does the nonextensive entropy $s$ of a
relatively small number of atoms of space combine into the extensive
Bekenstein--Hawking entropy $S_\mathsf{BH}$ of a macroscopic black hole?
Somehow, this may involve a form of collective behavior of a subset of the
atoms, ``monatomic molecules,'' perhaps even collective behavior of
combinations of different types of atoms, ``hetero-atomic molecules.'' (For a
related discussion of entropic modifications of Newton's law in the Verlinde
framework, see, e.g., Refs.~\refcite{ModestoRandono2010,Nicolini2010} and
references therein.)

Elaborating on the discussion of the previous paragraph,
there may be special mixtures of different types of atoms, which give
an effective index $q_\mathsf{eff}=1$ for the entropy
but an effective factor $I_{1,\,\mathsf{eff}}\ne 1$ for the
energy equipartition law. This may not be altogether impossible, as the
following simple argument shows. Imagine an equal mixture of two hypothetical
types of noninteracting atoms, $a$ and $b$, with $q_a=1-\Delta q$ and
$q_b=1+\Delta q$, for $0<|\Delta q|< 1$, and $k_{qa}=k_{qb}=k_{B}$.
Mathematically, there is $(1-q_{a})+(1-q_{b})$ $=$ $0$ and
$1/(2-q_{a})+1/(2-q_{b})$ $=$ $2/(1-(\Delta q)^2)$ $\ne$ $1+1$. Using a
short-hand notation and setting $k_{B}=1$, two independent systems,
$L$ and $R$, each with approximately equal numbers
of $a$-- and $b$--type atoms ($N_{La}\sim N_{Lb}$, $s_{La}\sim s_{Lb}$,
and similarly for $R$), then have the following total
entropy from \eqref{eq:s-nonextensive} and total energy from
\eqref{eq:E-equipartition} and \eqref{eq:I1-2dMaxwell}:
\begin{subequations}\label{eq:sLR-ELR}
\begin{eqnarray}
\hspace*{-2.0mm} s_{L+R} &=& s_{La}+s_{Ra}+ \big(1-q_a\big)\,s_{La}\,s_{Ra}+
s_{Lb}+s_{Rb}+ \big(1-q_b\big)\,s_{Lb}\,s_{Rb}
\nonumber\\[1mm]
\hspace*{-2.0mm} &\sim& s_{La}+s_{Lb}+s_{Ra}+s_{Rb} +
\big(1-q_a+1-q_b\big)\,s_{La}\,s_{Rb}
\nonumber\\[1mm]
\hspace*{-2.0mm}
&\sim& s_{La}+s_{Lb}+s_{Ra}+s_{Rb}
\nonumber\\[1mm]
\hspace*{-2.0mm}
&\sim& s_{L}+s_{R}\,,
\label{eq:sLR}\\[2mm]
\hspace*{-2.0mm} E_{L+R}/T &=&
\textstyle{\frac{1}{2}}\,\big(N_{La}/(2-q_a)+N_{Lb}/(2-q_b)\big)+
\textstyle{\frac{1}{2}}\,\big(N_{Ra}/(2-q_a)+N_{Rb}/(2-q_b)\big)
\nonumber\\[1mm]
\hspace*{-2.0mm} &\sim&
 N_{La}/\big(1-(\Delta q)^2\big)+
 N_{Rb}/\big(1-(\Delta q)^2\big)
\nonumber\\[1mm]
\hspace*{-2.0mm} &\sim&
\textstyle{\frac{1}{2}}\,
\big(N_{La}+N_{Lb}+N_{Ra}+N_{Rb}\big) \,1/ \big(1-(\Delta q)^2\big)\,.
\label{eq:ELR}
\end{eqnarray}
\end{subequations}
The above results effectively show
a standard (extensive) entropy and a modified energy
equi\-par\-ti\-tion law, at least, for the hypothetical types of atoms
of this simple argument. It remains to be seen if a similar result
holds for a mixture of atoms from Table~\ref{table1}
or atoms obtained from a more advanced calculation.

\section{Conclusion}\label{sec:Conclusion}

It may be helpful to give a brief summary of what has been achieved
in this paper. Consider, for
simplicity, the case of ``bits`` building up the macroscopic surface
(black-hole horizon), each bit contributing a quantum of area $l^2$ and
having an internal (Hilbert-space) dimension $d_\mathsf{atom}=\widehat{d}=2$.
The main result, then, is the simple formula \eqref{eq:G-final}
for Newton's gravitational coupling constant $G$, with $\widehat{d}=2$ for
the case of bits.

One possible physical interpretation uses the framework of
Verlinde.\cite{Verlinde2010}
In that framework,\cite{Verlinde2010,'tHooft1999}
a finite-temperature system
of bits on a holographic screen gives rise to Newtonian gravity
\eqref{eq:Agrav-derivation} with the above-mentioned coupling constant $G$.
However, in order to obtain an integer internal dimension $\widehat{d}$,
these bits must obey a nonstandard energy equipartition law
\eqref{eq:E-equipartition}, which may perhaps trace back to a type of
nonstandard statistics.\cite{Tsallis1988}
Most likely, the origin of this nonstandard behavior is some form of
long-range interaction between the bits themselves.

\section*{Acknowledgments}

The author thanks J.A.~Neto for a copy of Ref.~\refcite{Tsallis1999}
and H. Sahlmann and S. Thambyahpillai for useful comments.


\end{document}